\documentclass[preprint,showkeys,secnumarabic,amsfonts,amsmath,amssymb]{revtex4}
\usepackage[dvips]{color}
\usepackage{array}
\usepackage{epsfig}
\usepackage{amsmath}
\usepackage{graphicx}
\begin{document}
\title{Dynamical System Analysis of Brane Induced Gravity with Tachyon Field}

\author{A. Ravanpak}
\email{a.ravanpak@vru.ac.ir}
\affiliation{Department of Physics, Vali-e-Asr University of Rafsanjan, Rafsanjan, Iran}
\author{G. F. Fadakar}
\email{g.farpour@vru.ac.ir}
\affiliation{Department of Physics, Vali-e-Asr University of Rafsanjan, Rafsanjan, Iran}

\date{\today}

\begin{abstract}

In this manuscript we use the dynamical system approach to study the linear dynamics of a normal DGP brane-world model with a tachyon field as the dark energy component. Our focus is on a Gaussian tachyonic potential in which the parameter $\lambda=-V_\phi/V^{3/2}$, goes to infinity. One of the most important results of this study is that we find critical submanifolds which indicate the effect of extra dimension.

\end{abstract}


\maketitle

\section{Introduction}\label{s:1}

According to the big bang theory which is the prevailing cosmological model, the universe started its evolution and expansion from a state of extremely high temperature and density. In the very early stages, it experienced a very rapidly accelerated expansion phase within a very short period of time, called inflation. Then, it underwent a radiation dominated era, passed through a matter dominated regime and came into another accelerated expansion phase until now, dubbed dark energy (DE) dominated era. Many authors have investigated different stages of the history of the universe, separately. For example, there are many articles about various inflationary models \cite{Guth}-\cite{Setare}. Also, different DE scenarios have been proposed to explain the late time acceleration \cite{Caldwell}-\cite{Gao}. Maybe the most usual theoretical instrument to explain the inflationary era and also the most common candidate for DE, is the concept of scalar field. Among different kinds of scalar fields, tachyon field which originates from string theory is of particular interest. It has frequently been shown that the tachyon field can play the role of the inflaton field, and/or the role of the dark sectors of the universe \cite{Leblond}-\cite{Padmanabhan2}. It has a positive potential which has a maximum at $\phi=0$, and approaches zero when $\phi\rightarrow\infty$, whilst during the entire of this process, the slope of the potential is negative.

In addition to investigating different periods of the evolution of the universe distinctly, it could be very useful to study its whole history, all at once. Dynamical system approach is a well known procedure in this context which provides a mathematical tool for qualitative study of the behavior of complex dynamical systems, usually by employing ordinary differential equations \cite{Bahamonde}-\cite{Campos2}. It has its origins in Newtonian mechanics in which we try to obtain the trajectory of the system. But, the behavior of complex dynamical systems are too complicated to be understood in terms of these individual trajectories, because it may not be possible for instance to know all of the precise values of model parameters. Also, in many situations it is more important to know the type of trajectory than one particular trajectory. Dynamical system theory deals with the long-term qualitative behavior of dynamical systems. This qualitative study is based on stability analysis. The stability of dynamical systems implies the equivalent trajectories related to a class of models or initial conditions and classify all possible trajectories. Several notions of stability have been introduced in dynamical system approach.

On the other hand, the idea of higher dimensional theories of gravity has attracted a great deal of attention. Although it was mentioned first in Kaluza-Klein theory but it revived after the advent of string theory in Arkani-Hamed, Dimopoulos and Dvali scenario and Randall and Sundrum (RS) models \cite{Arkani}-\cite{Randall2}. Another interesting 5D scenario, proposed by Dvali, Gabadadze and Porrati (DGP), is a brane-induced gravity model which considers a 4D brane, that is embedded into
a 5D minkowski bulk \cite{Dvali}. This is a phenomenological model that could be considered as an inspiration of string theory. In this model, according to how the 4D brane embeds in the bulk, we obtain two separate branches which are distinguished with a parameter $\epsilon = \pm1$. The branch with $\epsilon=+1$, is called the self-accelerating branch because it results the late time acceleration of the universe naturally, without any need to have a DE component. But the other, $\epsilon=-1$, is the normal branch which needs a DE component to explain the late time acceleration.

In this manuscript, we will concentrate on the stability analysis of the tachyon DGP model assuming that the tachyon potential has a Gaussian form. The choice of tachyon field as a dark energy term on the brane in DGP cosmology has been studied in several papers \cite{Ravanpak2},\cite{Ping}-\cite{Zhang}. In \cite{Zhang}, the authors have investigated the stability of this combination model for two kinds of tachyonic potential: the exponential potential and inverse square potential and they didn't find any attractor critical point analytically, whereas in another article the authors have studied the stability of a DGP model with a quintessence scalar field and obtained the model's attractor critical points \cite{Quiros}. They have utilized both a constant potential and an exponential potential and especially revealed that there is an attractor submanifold in the case of a constant potential which indicates the effect of extra dimension. Here, using a Gaussian potential we find attractor critical points as well as such attractor submanifold that shows the effect of extra dimension. Also we describe evolution of the universe according to the critical points.

The paper is organized as follows: in Sec.\ref{s:2}, we review the basic equations of the model. In Sec.\ref{s:3}, We rewrite the equations of motion in terms of
autonomous differential equations for which we identify the critical points and analyze their stability condition for a inverse square and Gaussian
form of the potentials. Finally in Sec.\ref{s:4}, we present a summary and discuss our results.

\section{THE MODEL}\label{s:2}

The spatially flat Friedmann equation on the brane in the normal branch of DGP scenario considering tachyon field as the DE component is as follows \cite{Deffayet}
\begin{equation}\label{fried}
H^2+\frac{H}{r_c}=\frac{1}{3M_p^2}(\rho_m+\rho_{tac})
\end{equation}
Here $r_c$, which is the relation between the 4D and 5D Planck mass is called the crossover scale, $H$, is the Hubble parameter, $\rho_m$, is the matter energy density, $\rho_{tac}$, denotes the energy density of the tachyon field and $M_p$, is the 4D Planck mass. On the other hand, the energy density and pressure of tachyon field are expressed as
\begin{eqnarray}
  \rho_{tac} &=& \frac{V(\phi)}{\sqrt{1-\dot\phi^2}} \label{rho} \\
 P_{tac} &=& -V(\phi)\sqrt{1-\dot\phi^2} \label{p}
\end{eqnarray}
in which $\phi$ and $V(\phi)$ are the tachyon field trapped on the brane and the tachyon potential, respectively and dot means derivative with respect to the cosmic time. In the absence of any interaction between the dark sectors of the universe they satisfy the conservation equations as below
\begin{eqnarray}
  \dot\rho_m &+& 3H\rho_m = 0 \label{conservationdm} \\
  \dot\rho_{tac} &+& 3H(\rho_{tac}+P_{tac})= 0 \label{conservationtac}
\end{eqnarray}
where we have assumed the matter content of the universe as a perfect fluid with vanishing pressure, called dust. Replacing $\rho_{tac}$ and $P_{tac}$, in Eq.(\ref{conservationtac}), with Eqs.(\ref{rho}) and (\ref{p}), one can obtain the equation of motion of the tachyon field as
\begin{equation}\label{field}
\frac{\ddot\phi}{1-\dot\phi^2}+3H\dot\phi+\frac{V_\phi}{V}=0
\end{equation}
in which $V_{\phi}$, represents the derivative of the $V(\phi)$, with respect to the tachyon scalar field.

\section{phase space and stability analysis}\label{s:3}

As we mentioned in introduction we intend to describe the model as an autonomous system of ordinary differential equations. So, we define a new set of dimensionless dynamical variables
\begin{equation}\label{nv}
x^2=\frac{\rho_m}{3M_p^2(H^2+\frac{H}{r_c})}, \quad y^2=\frac{V}{3M_p^2(H^2+\frac{H}{r_c})}, \quad d=\dot\phi, \quad z^2=1+\frac{1}{Hr_c}, \quad \lambda=\frac{-M_pV_{\phi}}{V^{3/2}}
\end{equation}
Since $r_c$, is always positive, and in an expanding universe we have $H>0$, then $z\geq1$. On the other hand, in a contracting universe, i.e. $H<0$, we find $0\leq z\leq1$. So, the expanding universe and the contracting universe are independent submanifolds and one can study each of them, separately. Here, we focus on an expanding universe. The common subset $(x, y, d, z = 1)$, corresponds to the formal limit $r_c\rightarrow\infty$, which represents the standard behavior of 4D Einstein-Hilbert theory coupled to a tachyon field. Using the above new introduced variables, the Friedmann constraint, Eq.(\ref{fried}), reads
\begin{equation}\label{fried-cons}
x^2+\frac{y^2}{\sqrt{1-d^2}}=1
\end{equation}
while the Raychaudhury equation can be recast as
\begin{equation}\label{ray}
\frac{\dot H}{H^2}=\frac{-3z^2}{z^2+1}\left(x^2+y^2\frac{d^2}{\sqrt{1-d^2}}\right)
\end{equation}
Since $d$, appears under the square root and to have a physical meaning it should satisfy the constraint $-1\leq d\leq1$. Also, we can rewrite the first relation in Eq.(\ref{nv}), as \textbf{$x^2z^2=\Omega_m$}, in which $\Omega_m$, is the dimensionless density parameter of the matter content of the universe which changes between $0$ and $1$. Thus, using the constraint of $z$, we find that \textbf{$-1\leq x\leq1$}. From the Friedmann constraint and with attention to the range of changes of $x$, one can conclude that $-1\leq y\leq1$.

On the other hand, we can obtain the tachyon EoS parameter and the total EoS parameter of the universe in terms of the dimensionless variables as
\begin{eqnarray}
    w_{tac} &=& d^2-1 \\\label{w}
    w_{tot} &=& -y^2\sqrt{1-d^2}\label{wtot}
\end{eqnarray}
The set of evolution equations of the model under consideration can be obtained using the Eqs.(\ref{nv}), (\ref{fried-cons}) and (\ref{ray}), as below
\begin{eqnarray}
  d' &=& -(3d-\sqrt{3}zy\lambda)(1-d^2) \\\label{e1}
  y' &=& -\frac{\sqrt{3}}{2}y^2dz\lambda+\frac32y(1-y^2\sqrt{1-d^2}) \\\label{e2}
  z' &=& \frac32\frac{z(z^2-1)}{z^2+1} (1-y^2\sqrt{1-d^2}) \\\label{e3}
  \lambda' &=& -\sqrt{3}\lambda^2dyz(\Gamma-3/2)\label{e4}
\end{eqnarray}
Here, prime means derivative with respect to $\ln a$, and $\Gamma=VV_{\phi\phi}/V_{\phi}^2$, where by $V_{\phi\phi}$, we mean the second derivative of the potential with respect to the tachyon field. Also, we have removed the new variable $x$, in the above equations using the Friedmann constraint. These equations form a four dimensional autonomous system and indicate the evolution of the phase space variables $d$, $y$, $z$ and $\lambda$, and indirectly the behavior of the DGP model with tachyon field.

In stability formalism, we try to solve $d'=y'=z'=\lambda'=0$, simultaneously, and find the critical points of the model and study their existence conditions using respective eigenvalues. The form of the potential $V(\phi)$, plays an important role in this scenario. It is more convenient henceforth to divide our discussion into two parts depending on the parameter $\lambda$. We consider two general cases:

\begin{description}
  \item[1.] $\lambda=$ constant. With attention to the definition of $\lambda$, this choice yields an inverse square potential, $V(\phi)\propto\phi^{-2}$.
  \item[2.] $\lambda=\lambda(\phi)$. Various potentials lead to a varying $\lambda$. But, in the following we are going to consider a Gaussian potential, $V(\phi)\propto e^{-\gamma\phi^2}$ which has a maximum value
at $\phi=0$ and decays to zero as $\phi$ goes to infinity that is suitable for a tachyonic potential.
\end{description}

\subsection{The case $\lambda=$ constant}

Obviously, the case $\lambda=$ constant, relates to $\lambda'=0$. In this situation when we integrate the equation $\lambda=-V_\phi/V^{3/2}$, we conclude that the tachyon potential must have an inverse square form as $V(\phi)=V_0\phi^{-2}$. This result is just like the one in \cite{Aguirregabiria}-\cite{Nozari}. Also, in \cite{Mizuno}, the authors obtained such a potential for a quintessence scalar field in the early stages of a brane-world RSII model. It is easy to show that in this case $\Gamma=3/2$, which from Eq.(\ref{e4}), results $\lambda'=0$. The fixed points of the system can be obtained by setting $d'=y'=z'=0$. The results have been given in TABLE \ref{table:1}, in which $y^{\ast}=\sqrt{\alpha/6}$, $d^{\ast}=\lambda y^{\ast}/\sqrt3$, $\alpha=\sqrt{\lambda^4+36}-\lambda^2$ and $\delta=-3+\alpha\lambda^2/72$. We must note that we have found some other critical points in our model, but we have not mentioned them in TABLE \ref{table:1}, because either they did violate the Friedmann constraint or they did not satisfy the constraints on the dynamical variables.

\begin{table}[h]
\caption{The fixed points of the model for $\lambda=$ constant.} 
\centering
\begin{tabular}{|c|c|c|c|c|c|}
  \hline
  point  \ & $(d,y,z)$ \ & eigenvalues\ & $w_{tot}$ \ & stability \ & solution\\ [1ex] 
\hline\hline
  $P_1$ & $(0,0,1)$ & $(-3,3/2,3/2)$  & 0 & saddle & matter dominated\\
  \hline
  $P_2^\pm$ & ($\pm d^{\ast}$, $\pm y^{\ast}$,1) & $(-6(3+\delta),\delta,\delta)$ & $-\frac{\alpha}{6}\sqrt{1-\frac{\alpha\lambda^2}{18}}$ & stable & tachyon dominated \\
  \hline
   $P_3^\pm$ & ($\pm 1$,0,1) & $(6,3/2,3/2)$ & 0 & unstable & matter scaling\\
  \hline
\end{tabular}
\label{table:1} 
\end{table}\

The critical point $P_1$ is a saddle point, because one of its eigenvalues is negative and the others are positive. Since from Friedmann constraint we find $x^2=1$, using the relation $\Omega_m=x^2z^2$, we conclude that this point refers to a matter dominated solution, $\Omega_m=1$. One can see in TABLE \ref{table:1} that at this critical point $w_{tot}=0$, which is in agreement with a matter dominated regime. The critical points, $P_3^\pm$, with all positive eigenvalues represent an unstable critical point. Although the universe behaves like a matter dominated era, because $w_{tot}=0$, but against $P_1$, the kinetic term of the tachyon field has a remarkable contribution at these points. So, we can call them matter scaling solutions. To specify the characteristics of critical points $P_2^\pm$, we need to investigate the parameter $\lambda$, more carefully. Using its definition and considering the negative slope of a tachyonic potential, one can find that $\lambda>0$. It cannot be zero because $\lambda=0$, means $V_\phi=0$, which for an inverse square potential, results $V_0=0$ and consequently $V=0$. Using the constraint $0<\lambda<\infty$, and the definition of $\alpha$, one can obtain another constraint as $0<\alpha<6$. It is easy to check that for each pair of values of $(\lambda,\alpha)$, $\alpha\lambda^2\leq18$. This also can be deduced from the relation of $w_{tot}$. Thus, we obtain an interval for the parameter $\delta$, as $-3<\delta<-11/4$. So, the eigenvalues respective to the critical points $P_2^\pm$, are always negative and therefore they are attractor solutions. Also, from Eq.(\ref{nv}), we can write $\Omega_{tac}=\frac{\rho_{tac}}{3M_p^2H^2}=\frac{y^2z^2}{\sqrt{1-d^2}}$. One can check that $\Omega_{tac}=1$, at points $P_2^\pm$. Therefore, we call them tachyon dominated solutions. On the other hand, from $w_{tot}<-1/3$, we reach the condition of acceleration as $\alpha\sqrt{1-\frac{\alpha\lambda^2}{18}}>2$. So, one can check that for $0<\lambda<1.86$, these critical points show a de Sitter solution, while for $\lambda\gtrsim1.86$, they are just tachyon dominated solutions, without acceleration. This result is similar to the case reported in \cite{Copeland2} and \cite{Gumjudpai}.

In addition to what we mentioned above about these critical points, their similarity is that all of them belong to a 4D universe, because in all of them $z=1$. Thus, extra dimension has no significant effect in the case $\lambda=$ constant. Trajectories may exit the plane $z=1$, in parts of the history of the universe, but they must return to the 4D universe as these critical points show.

\subsection{The case $\lambda=\lambda(\phi)$}

When the potential form is something different from the inverse square, $\lambda$, will be a dynamically changing quantity. In the following we choose a Gaussian potential. Assuming $\lambda$, evolves sufficiently slow such that one can take it to be a constant within a short period of the evolution of the universe, we can regard the critical points in the previous subsection for the constant $\lambda$, as the instantaneous critical points of the current dynamical system \cite{Copeland}-\cite{Ng}. This interesting assumption is useful in order to see where the solution tends to at that instant, because according to this assumption, $P_2^\pm$, are dynamical critical points. Also, in addition to the fixed points $P_1$ and $P_3^\pm$, and the moving critical points $P_2^\pm$, which exhibit a standard 4D behavior because of $z=1$, we find two critical submanifolds in this situation, $P_4^\pm: (d=0, y=\pm1, z)$, that show the effect of extra dimension and only exist for $\lambda=0$, that corresponds to extremum of the Gaussian potential where $V_{\phi}=0$. In this case one can check that $\Omega_{tac}=1$. On the other hand, using Eq.(\ref{wtot}), we find $w_{tot}=-1$. Hence, we conclude that these submanifolds indicate a tachyon dominated de Sitter solution.

Further, we obtain the respective eigenvalues for $P_4^\pm$ submanifolds, as $(0,-3,-3)$. Since at least there is one zero eigenvalue, we cannot use the linear approximation method to discuss their stability, rather we turn to a method based on center manifold theory \cite{Wiggins}-\cite{Fang}. In this scenario, one has to reduce the dimensionality of the dynamical system under consideration and then, the stability of this reduced system is studied instead of the main one. The interesting feature of this method is that the stability of the critical points of the main system can be understood via the stability properties of this reduced system. The mathematical procedure is complicated and out of scope of the current work, but concisely one must calculate the center manifold related to that critical point and investigate the dynamics on the center manifold to comment on the stability of that critical point. But, in the case of a critical line, with only one zero eigenvalue, such as the case here, the respective center manifold is nothing but the critical line itself \cite{Bahamonde}. Therefore, in the case of $P_4^\pm$, the $z$-axis is the center manifold, and we can expect that orbits near the $z$-axis are perpendicularly attracted or repelled
according to the sign of the non vanishing eigenvalues. Since the non vanishing eigenvalues are all negative, then we conclude that $P_4^\pm$, are attractors.

Since $\lambda$, is a varying quantity, its asymptotic behavior has a crucial role in a complete understanding of the nature of the dynamical attractor critical points $P_2^\pm$. To this aim, we return to Eq.(\ref{e4}). As we mentioned above, the case $\lambda=$ constant, relates to $\Gamma=3/2$. Also, it is generally easy to show that for positive values of $d$, and $y$, if $\Gamma>3/2$, $\lambda\rightarrow0$, and if $\Gamma<3/2$, $\lambda\rightarrow\infty$, asymptotically. But, for a common Gaussian potential in the form $V=V_0e^{-\gamma\phi^2}$, we find that $\Gamma=1-1/(2\gamma\phi^2)$. So, the maximum value of $\Gamma$, that is achieved in the limit $\phi\rightarrow\infty$, is $\Gamma=1$. Therefore, for a Gaussian potential, $\lambda\rightarrow\infty$, asymptotically. Also, according to discussions in the prior subsection, when $\lambda$, increases from zero to infinity, $\alpha$, decreases from 6 to zero, so that the relation $\alpha\lambda^2\leq18$, is always satisfied. In fact, when $\lambda=0$, $\alpha=6$, and $P_2^{\pm}$ start from $(0,\pm 1,1)$ on the $P_4^{\pm}$, with $w_{tot}=-1$, and when $\lambda\rightarrow\infty$, $\alpha\lambda^2\rightarrow18$, and consequently $P_2^{\pm}$ approach $P_3^{\pm}$, i.e., $y^{\ast}\approx\sqrt3/\lambda\rightarrow0$, and $d^{\ast}\rightarrow1$, while they are still attractor solutions with $w_{tot}=0$. Thus, with attention to descriptions above, it seems that the universe will experience a phase transition from acceleration to deceleration during the evolution of $\lambda$. But in fact, it depends crucially on how fast the system reaches a neighborhood of $P_2^{\pm}$. If the universe reaches $P_2^{\pm}$, while $0<\lambda<1.86$, one can conclude that it will experience a temporal acceleration before entering a deceleration phase. Otherwise, it will never come across an accelerating phase. If we correspond the current acceleration of our universe to the one we mentioned above, it will eventually enter a decelerating phase. \cite{Copeland}-\cite{Copeland2}.

\begin{figure*}
\centering
\includegraphics[width=7cm]{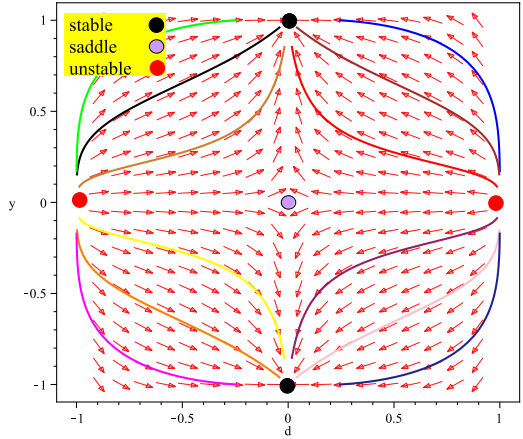}
\caption{The 2D phase space $(y,d)$, for $\lambda=0$. The universe starts from matter scaling solution and tends to reach tachyon dominated critical point.}\label{fig1}
\end{figure*}

Fig.\ref{fig1}, shows the 2D phase plane $(y,d)$ of the model under consideration for $\lambda=0$. The universe starts from the unstable matter dominated critical point and tends to reach the stable dark energy dominated critical point. The important property for the case $\lambda=0$, is the appearance of $P_4^\pm$ submanifolds.

\begin{figure*}
\centering
\includegraphics[width=7cm]{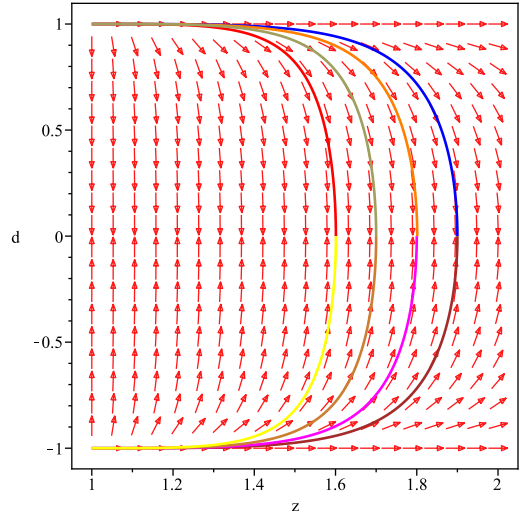}
\caption{The 2D phase space $(d,z)$, for $\lambda=0$ and $y=\pm1$. The line $d=0$, refers to $P_4^\pm$ submanifolds.}\label{fig2}
\end{figure*}

Apart from the discussion above in utilizing the center manifold theory in investigating the stability of the critical line $P_4^{\pm}$, we use numerical techniques to show the behavior of trajectories in the vicinity of this critical line. Fig.\ref{fig2}, indicates the 2D phase plane $(d,z)$ of the model for $y=\pm1$. The line $d=0$, shows these stable critical submanifolds in this 2D portrait.

\begin{figure*}
\centering
\includegraphics[width=5cm]{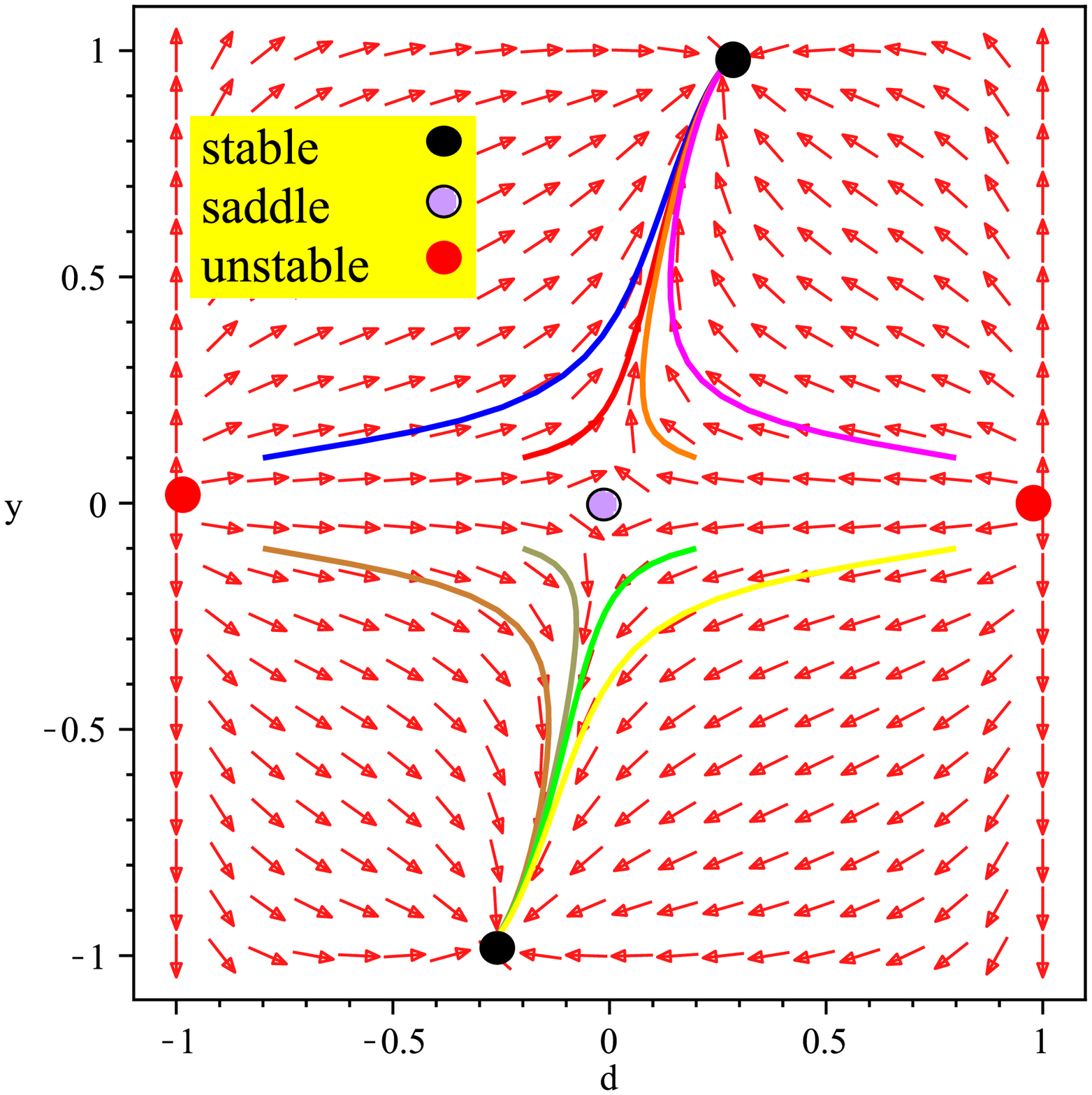}
\includegraphics[width=5cm]{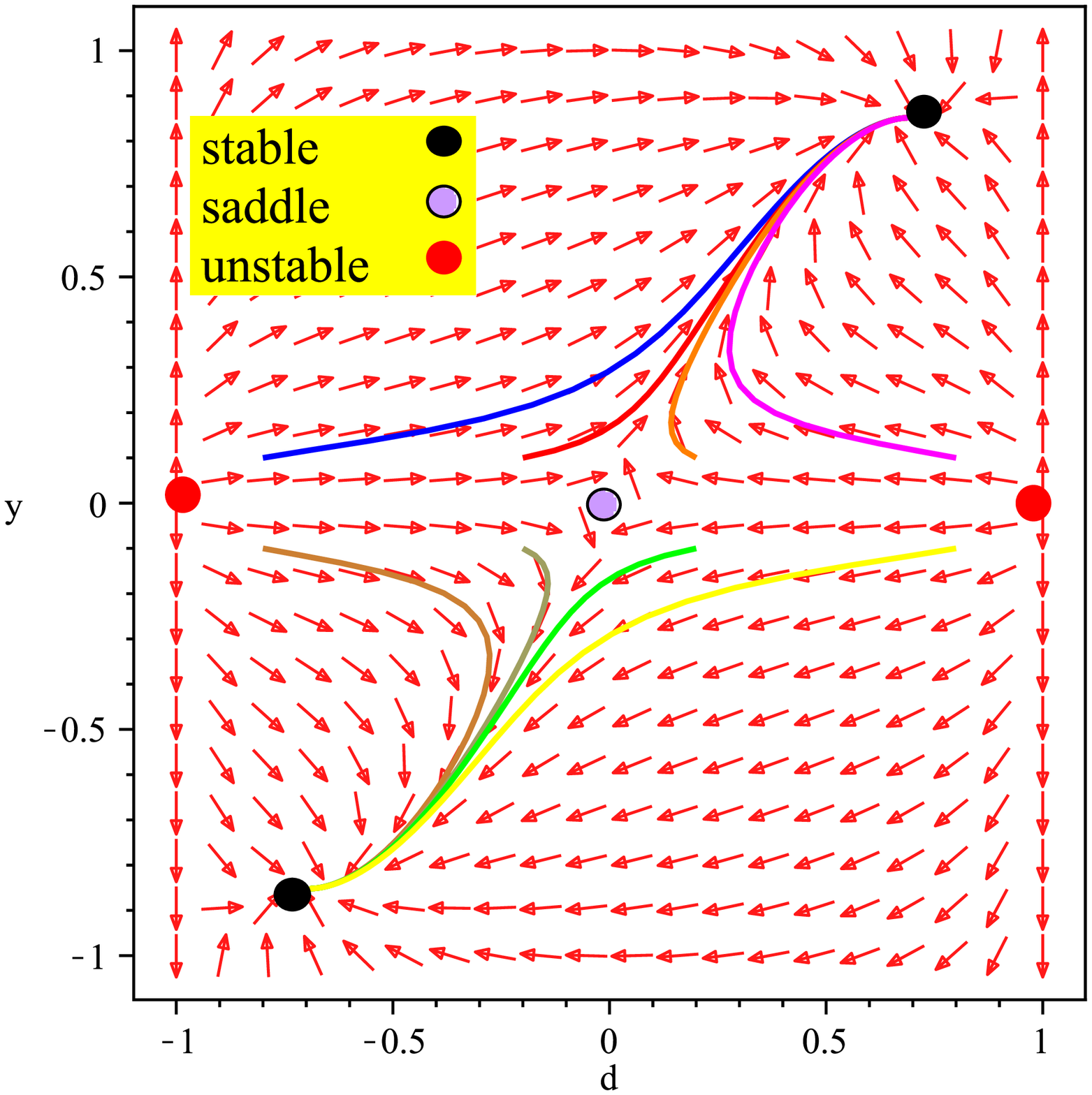}
\includegraphics[width=5cm]{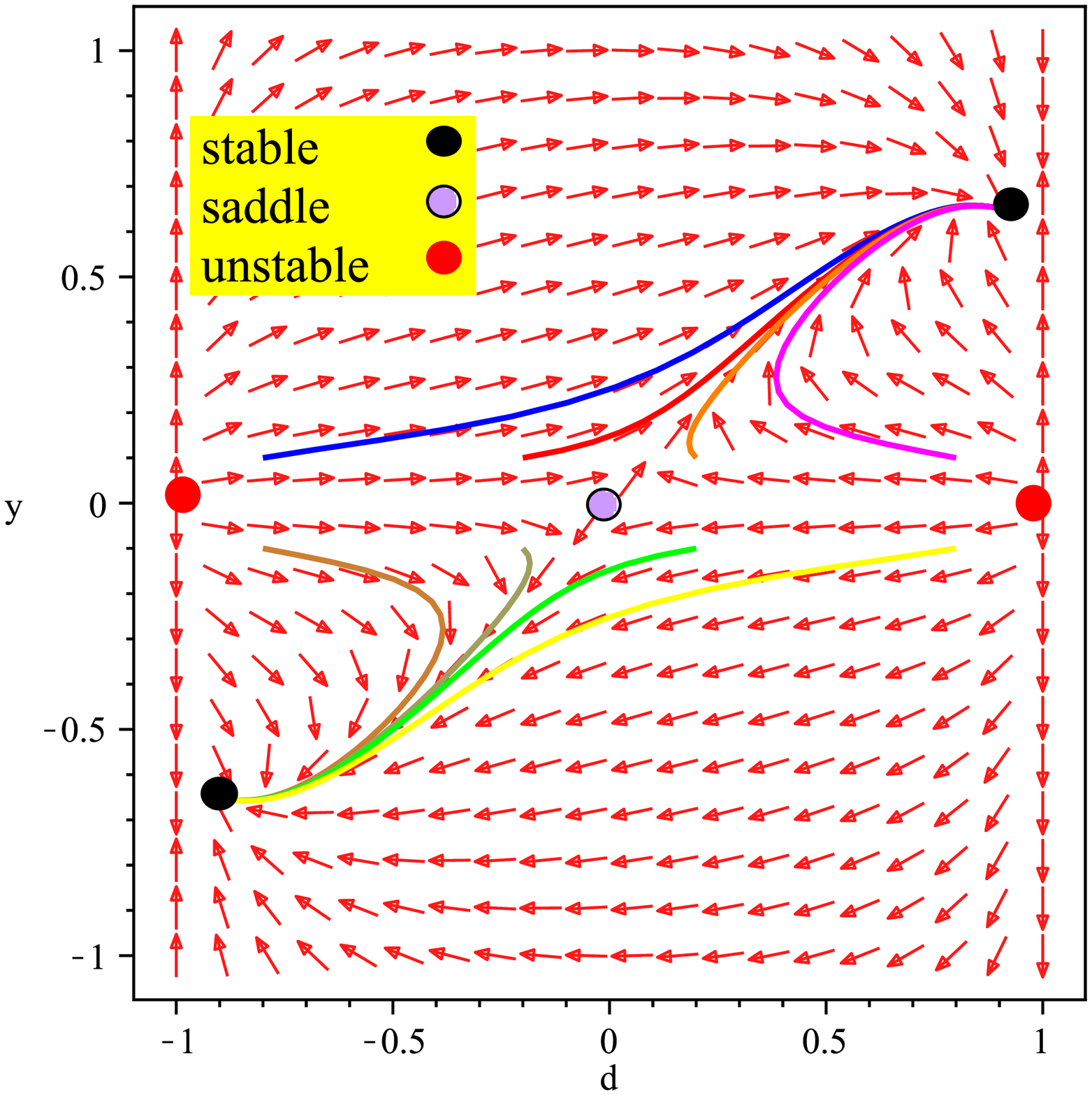}
\caption{The 2D phase space $(y,d)$, for different values of $\lambda$. From left to right: $\lambda=0.1$, $\lambda=1.4$ and $\lambda=2.4$. Attractor solutions $P_2^\pm$, are instantaneous critical points. They approach $P_3^\pm$ and during this process the universe experiences a transition from acceleration to deceleration.}\label{fig3}
\end{figure*}

Fig.\ref{fig3}, consists of three 2D phase portrait $(y,d)$ of our dynamical system for different values of parameter $\lambda$. It is obvious from these figures that along with increasing the value of $\lambda$, the stable critical points $P_2^\pm$, move around and approach $P_3^\pm$, while they are still attractors and as well as while universe has experienced a transition from acceleration to deceleration during this process.

\section{Conclusion}\label{s:4}

In the present work we studied a DGP brane-world model with a tachyon scalar field on the brane for two different kinds of tachyon potential. Our main focus were on the Gaussian potential for which $\lambda$, is not constant. Then, we followed the dynamical system approach to understand the evolution of the universe in this model. One of our most important results was that we found stable critical submanifolds that depend on the new variable $z$, which relate to the extra dimension. We should note that in a tachyon DGP model, these attractor submanifolds only exist for those tachyonic potentials that have an extremum, such as the Gaussian potential in the present work. On the other hand, we found two dynamical critical points $P_2^\pm$ that assuming slowly moving, one can consider them as instantaneous fixed critical points. The most interesting feature of this assumption is that we can study the behavior of the universe for different values of $\lambda$. Also, we found that with the expansion of the universe, as $\phi$ goes to infinity, the tachyonic potential will decrease to zero and critical points $P_2^\pm$, approach $P_3^\pm$ while they remain attractors. During this evolution, the behavior of the total equation of state parameter suggested that the universe may experience a transition from acceleration to deceleration. Future observations may show if such a transition is possible or not.

\acknowledgments The authors would like to thank Nelson J. Nunes for his valuable and helpful comments.

\nocite{*}
\bibliographystyle{spr-mp-nameyear-cnd}

\end{document}